\documentclass[12pt,twoside]{article}

\usepackage{amssymb}

\date{}

\begin{document}

\centerline{\bf Applied Mathematical Sciences, Vol. 1, 2007, no. 17,
843 - 852}

\centerline{}

\centerline{}

\centerline {\Large{\bf FRW BAROTROPIC ZERO MODES:}}

\centerline{}

\centerline{\Large{\bf DYNAMICAL SYSTEMS OBSERVABILITY}}

\centerline{}

\centerline{\bf {H.C. Rosu}}

\centerline{}

\centerline{Instituto Potosino de Investigaci\'on Cient\'{\i}fica y
Tecnol\'ogica}

\centerline{Apdo Postal 3-74 Tangamanga, 78231 San Luis Potos\'{\i},
S.L.P., Mexico}

%\centerline{Address of Author1 third line}

%\centerline{Address of Author1 forth line}

\centerline{}

\centerline{\bf {V. Ibarra-Junquera}}

\centerline{}

\centerline{Faculty of Chemical Sciences, University of Colima}

\centerline{28400 Coquimatl\'an, Col., Mexico}

%\centerline{Address of Author2 third line}

%\centerline{Address of Author2 forth line}

\newtheorem{Theorem}{\quad Theorem}[section]

\newtheorem{Definition}[Theorem]{\quad Definition}

\newtheorem{Corollary}[Theorem]{\quad Corollary}

\newtheorem{Lemma}[Theorem]{\quad Lemma}

\newtheorem{Example}[Theorem]{\quad Example}

\begin{abstract} %The paper must have abstract not exceeding 200 words.
The {\em dynamical systems observability} properties of barotropic
bosonic and fermionic FRW cosmological oscillators are investigated.
Nonlinear techniques for dynamical analysis have been recently
developed in many engineering areas but their application has not
been extended beyond their standard field. This paper is a small
contribution to an extension of this type of dynamical systems
analysis to FRW barotropic cosmologies. We find that determining the
Hubble parameter of barotropic FRW universes does not allow the {\em
observability}, i.e., the determination of neither the barotropic
FRW zero mode nor of its derivative as dynamical cosmological
states. Only knowing the latter ones correspond to a rigorous
dynamical observability in barotropic cosmology.

\end{abstract}



%\section{Introduction} This is the text of the introduction [1], [2].

\section{Introduction}

Over the years, modern dynamical systems theory has been applied
with considerable success to the evolution of cosmological models.
Many results concerning the possible asymptotic cosmological states
(at both early and late times) have been obtained and compared with
Hamiltonian methods and numerical studies in the authoritative book
of Wainwright and Ellis \cite{we97}.

Dynamical systems observability (DSO henceforth) is a rigorous
mathematical concept used to investigate if it is possible to know
the internal functioning of a given dynamical system. In the case of
cosmology, DSO can be considered a criterion of whether the possible
states of the dynamical universe as defined by some intelligent
observers can be indeed estimated. More precisely, one says that a
dynamical system is {\em observable} if it allows the usage of the
available information about the input $u(t)$ and the output $y(t)$
of the system to estimate the states $X(t)$ of the system. Thus, the
main idea behind (engineering) observability is to find the
dynamical states of a system based on the knowledge of its output.

In this paper we will investigate a cosmological application of this
type of observability analysis. In the context of FRW barotropic
cosmologies one important measured output is the Hubble parameter
and an interesting problem is if we can determine the so-called
barotropic zero modes since these modes can be identified as the
normal dynamical cosmological states. In the conformal time
variable, the logderivative of these modes determine the Hubble
parameter and thus they are equivalent to the comoving time scale
factors. The answer that we get according to the (engineering)
observability analysis is that we cannot determine the FRW zero
modes and therefore the scale factor if the measurable output is the
conformal Hubble parameter. However, knowing the latter ones or
their derivatives leads to a rigorous dynamical observability in
barotropic cosmology. We choose to present in the first sections of
the paper the physical part of the problem and the direct
mathematical application, whereas in the last section we collect the
required mathematical results. We end up with a short conclusion
section.

%\section{FRW barotropic zero modes}

\medskip

\noindent
%The barotropic FRW cosmologies obey the following set of differential equations:
%The scale factor ${\rm a(t)}$ of a FRW metric is a function of the

\section{FRW barotropic cosmology}

Barotropic cosmological zero modes are simple trigonometric and/or
hyperbolic solutions of second order differential equations of the
oscillator type to which the FRW system of equations can be reduced
when it is passed to the conformal time variable. They have been
discussed in a pedagogical way by Faraoni \cite{far} and also were
the subject of several recent papers \cite{zeromodes}. In this
section, we briefly review their mathematical scheme.

Barotropic FRW cosmologies in comoving time ${\rm t}$ obey the
Einstein-Friedmann dynamical equations for the scale factor ${\rm
a(t)}$ of the universe supplemented by the (barotropic) equation of
state of the cosmological fluid
%................................................
$$
%\left\{
\begin{array} {lll}
\rm  \frac{\ddot{a}}{a}=-\frac{4\pi G}{3}(\rho +3p)~, \\
%\eqno(1)
%\end{equation}
%\begin{equation} \label{e2}
\rm  H_{0}^{2}(t)\equiv\left(\frac{\dot{a}}{a}\right)^2=\frac{8\pi G\rho}{3}-\frac{\kappa}{a^2}~,\\
%\eqno(2)
%\end{equation}
%\begin{equation} \eqno{e3}
\rm p=(\gamma -1)\rho~,
%\eqno(3)
\end{array}
%\right.}
$$
%.................................................
where $\rho$ and ${\rm p}$ are the energy density and the pressure,
respectively, of the perfect fluid of which a classical universe is
usually assumed to be made of, $\kappa=0,\pm1$ is the curvature
index of the flat, closed, open universe, respectively, and $\gamma$
is the constant adiabatic index of the cosmological fluid.

%\medskip

%\underline{{\em Bosonic FRW barotropy}}

%\medskip

\subsection{FRW barotropic bosonic zero modes}

\noindent %Combining the equations (1)-(3) and using
Passing to the conformal time variable $\eta$, defined through
${\rm dt=a(\eta)d\eta}$, one can combine the three equations in a single Riccati equation for the Hubble parameter %one
%............................................
${\rm H_0}(\eta)$ (we shall use either $\rm \frac{d}{d \eta}$ or $'$
for the derivative with respect to $\eta$ in the following)
%..................................................
\begin{equation} \label{ricc}
{\rm H^{'}_{0}+cH^2_{0}+\kappa c=0~,}
%\eqno(6)
\end{equation}
%....................................................
where ${\rm c=\frac{3}{2}\gamma -1}$.

\medskip

Employing now ${\rm H_0(\eta)=\frac{1}{c}\frac{w^{'}}{w}}$ one gets
the very simple (harmonic oscillator) second order differential
equation
%.............................................
\begin{equation} \label{w}
{\rm w^{''}-c\cdot c_{\kappa,b}w=0~,}
%\eqno(7b)
\end{equation}
%............................................
where ${\rm  c_{\kappa,b}=-\kappa c}$. The solutions are the
following:
%\medskip
$$
{\rm For} \quad \kappa =1\, : \quad {\rm w_{1,b} \sim \cos(c\eta
+\phi)}\qquad \rightarrow \qquad {\rm w_{1,b}} \sim \left[{\rm
a}_{1,{\rm b}}(\eta)\right]^{{\rm c}}
%\sim {\rm w_{1}^{1/c}}  % [\cos(c\eta +d)]^{1/c}
~,
$$
where $\phi$ is an arbitrary phase. %whereas
$$
{\rm For} \quad \kappa =-1\, : \quad {\rm w_{-1,b}} \sim {\rm sinh
(c\eta)}\qquad \rightarrow \qquad {\rm w_{-1,{\rm b}}} \sim
\left[{\rm a_{-1,b}(\eta)}\right]^{\rm c}
%\sim {\rm w_{-1}^{1/c}}    %[{\rm sinh}(c\eta)]^{1/c}
~.
$$
%where ${\rm W_{\pm 1}}$ and ${\rm A_{\pm 1}}$ are amplitude parameters.
%Eqs. (8) and (9)
%According to Faraoni, the alternative method requires $\gamma$ be a
%constant.

Moreover,  the particular Riccati solutions
%${\rm u_{p}=\frac{1}{c}\frac{w^{'}}{w}}$ mentioned above is only the particular solution, i.e.,
${\rm H_{0}^{+}=-\tan c\eta}$ and ${\rm H_{0}^{-}={\rm coth}
\,c\eta}$ for $\kappa =\pm 1$, respectively, are related to the
common factorizations of the equation (\ref{w})
%....................................................BOSONIC FACTORIZATION
\begin{equation} \label{w3}
{\rm \left(\frac{d}{d\eta}+cH_{0}\right)
\left(\frac{d}{d\eta}-cH_{0}\right)w=
w^{''}-c(H_{0}^{'}+cH_{0}^{2})w=0}~.
%\eqno(7c)
\end{equation}
%.................................................
%......................................................
%For $\kappa =1$
Borrowing a terminology from supersymmetric quantum mechanics, we
call the solutions $\rm w$ as bosonic zero modes. As one can see,
they are actually some powers of the scale factors of the barotropic
universes.

%
%\bigskip

\subsection{FRW barotropic fermion zero modes}

%\medskip

\noindent A class of barotropic FRW cosmologies with inverse scale
factors with respect to the bosonic ones can be obtained  by
considering the supersymmetric partner (or fermionic) equation of
Eq.~(\ref{w3}) which is obtained by applying the factorization
brackets in reverse order
%...........................................
\begin{equation} \label{f}
{\rm \left(\frac{d}{d\eta}-cH_{0}\right)
\left(\frac{d}{d\eta}+cH_{0}\right)w=
w^{''}-c(-H_{0}^{'}+cH_{0}^2)w= 0}~.
\end{equation}
Thus, one can write
%...........................
\begin{equation}\label{f1}
{\rm w^{''} -c\cdot c_{\kappa, f}w=0}~,
%\eqno(7d)
\end{equation}
%.........................................
where
%..........................................
$$
{\rm c_{\kappa, f}(\eta)=-H_{0}^{'}+cH_{0}^2= \left\{
\begin{array}{ll}
c(1+2{\rm tan}^2 c\eta) & \mbox{if $\kappa =1$}\\
c(-1+2{\rm coth}^2 c\eta) & \mbox{if $\kappa =-1$}
\end{array} \right.}
%\eqno(7d)
$$
%...........................................
%\end{equation}
denotes the supersymmetric partner adiabatic index of fermionic type
associated through the mathematical scheme to the constant bosonic
index. Notice that the fermionic adiabatic index is time dependent.
The fermionic $\rm w$ solutions are
$$
{\rm w_{1,f} =\frac{c}{\cos (c\eta +\phi)}} \qquad \rightarrow
\qquad {\rm a_{1,f}(\eta) \sim [\cos(c\eta +\phi)]^{-1/c}}~,
$$
and
$$
{\rm w_{-1,f} =\frac{c}{sinh (c\eta)}}\qquad \rightarrow \qquad {\rm
a_{-1,f}(\eta) \sim [{\rm sinh}(c\eta)]^{-1/c}}~,
$$
for $\kappa =1$ and $\kappa =-1$, respectively.

\medskip

We can see that the bosonic and fermionic barotropic cosmologies are
reciprocal to each other, in the sense that
$$
{\rm a_{\pm,b}a_{\pm,f}=const}~.
$$
Thus, bosonic expansion corresponds to fermionic contraction and
viceversa.

%\bigskip

%\medskip

%**************************************************************************************
\noindent A matrix formulation of the previous results is possible
employing two Pauli matrices \cite{roj05}.
%%33333333333333333333333333333333333333333333333333333333333333333
The following spinor zero mode is obtained:
$$
{\rm W}=%\left( {\rm \begin{array}{cc}
%w_1\\
%w_2\end{array}} \right )=
\left( {\rm \begin{array}{cc}
{\rm w_{\kappa,f}}\\
{\rm w_{\kappa,b}}\end{array}} \right )~,
$$
showing that the two reciprocal barotropic zero modes enter on the
same footing as the two components of the spinor $\rm W$.

%************************************************************************************************************

\section{DSO Analysis of the FRW Barotropic Zero Modes}

The DSO analysis is based on the concept of observability space
$\mathcal{O}$, which is a smooth codistribution defined in an
algorithmic way by Isidori \cite{Isidori}. All the basic
mathematical material related to this concept is presented in
section 5 of this paper. %In this section, we write the conformal
%time oscillator equations in the dynamical systems form and
%introduce the corresponding nonlinear outputs for which we apply the
%algorithmic observability criterion of Isidori.

Frequently, the following form is used in controlled systems

\begin{eqnarray} \Gamma: \ \ \ \left \{
\begin{array}{l}
\dot{X}=\mathbf{f}(X)+\mathbf{g}(X)\mathbf{u} \label{afin en u}
\\\,\,y=h(X) \label{salida}
\end{array}
\right.
\end{eqnarray}

\noindent where $\mathbf{u} \in \Omega \subset R^{1}$, $X \in
\mathcal{S} \subset R^{2}$ and $\mathbf{f}$ and $\mathbf{g}$ are
$\mathcal{C}^\infty $ functions. Note that in general
$f(X)=\mathbf{f}(x)+\mathbf{g}(X)\mathbf{u}$. For the sake of
simplicity we denote by $\Gamma$ the set of equations \ref{afin en
u}.% and \ref{salida}.

In this section, we write the conformal time oscillator equations in
the above dynamical systems form and introduce the corresponding
nonlinear outputs for which we apply the algorithmic observability
criterion of Isidori according to the basic concepts in section 5.

%..................................................................................................THE BOSONIC CASE
\subsection{The Bosonic Case}
The dynamical system for this case is based on Eq.~(\ref{w})
%
%\begin{eqnarray}
%   F_1(t) &=& c \left( 1+2\, \left( \tan \left( c\ t \right)  \right) ^{2} \right) \\
%   F_2(t) &=& c \left( -1+2\, \left( \coth \left( c\ t \right)  \right) ^{2} \right) \\
%   c &=& \frac{{3}\ {\gamma}}{2}-1
%\end{eqnarray}
\begin{eqnarray}\Gamma _b: \ \ \ \left \{
\begin{array}{l}
    \dot{X}_1 = X_2\\
    \dot{X}_2 = -\ c\ (-\kappa \cdot c)\ X_1
    \end{array} \right.
\end{eqnarray}
In matrix form
\begin{eqnarray}
    \dot{X} = f_j(X)~,
\end{eqnarray}
where $X \in \mathbb{R}^2$, with $j=1$ when $\kappa=1$ and $j=2$
when $\kappa=-1$. %Assuming

Suppose that we take as outputs
\begin{equation}\label{outp1}
    h_1 = \frac{X_2}{X_1}\equiv \frac{\dot{X}_1}{X_1}\equiv {\rm c}\, {\rm H}_0(\eta) \qquad
    h_2 = \frac{X_1}{X_2}\equiv 1/h_1
    \end{equation}
or just
\begin{equation}\label{output2}
  h_3 = X_1\equiv w \qquad
  h_4 = X_2 \equiv w'~.
\end{equation}

Then, we can write the codistribution

\begin{eqnarray}
    \mathcal{O}_{j,i} &=& \left[ h_i(X), L_{f_j} h_i(X)\right]\\%^2\\
    \Xi_{j,i} &=& \verb"d"\ \mathcal{O}_{j,i}
\end{eqnarray}

This gives for $j=1$:
%
%\begin{eqnarray}
%\mathcal{O}_{b,1,1} &=&    \left[ \begin {array}{cc} -{\frac {X_{{2}}}{{X_{{1}}}^{2}}}&{X_{{1}}}^{-1}\\\noalign{\medskip}2\,{\frac {{X_{{2}}}^{2}}{{X_{{1}}}^{3}}}&-2
%\,{\frac {X_{{2}}}{{X_{{1}}}^{2}}}\end {array} \right]\nonumber
%\end{eqnarray}
%
\begin{equation}
   \Xi_{b,1,1}  = -\left[ \begin {array}{cc} {\left(-\frac{X_2}{X_1}\right)\frac {1}{{X_1}}}&\ \frac{1}{X_1}\\
    {\left(-\frac{X_2}{X_1}\right)\frac {2{X_2}}{{X_1}^2}}& {\frac {2 X_2}{{X_1}^{2}}}\end {array} \right]~, \nonumber
    \,\,\,
\Xi_{b,1,2} =     -\left[ \begin {array}{cc} \frac{1}{X_2} & \left(-\frac {X_1}{X_2}\right) \frac{1}{X_2} \\
  \frac {2{X_1}\ c^2\  } {{X_2}^2} & \left(-\frac {X_1}{X_2}\right)  \frac {2{X_1}\ c^2\  } {{X_2}^2}\end {array} \right]~. \nonumber
\end{equation}

Thus, for this case there exits a linear dependence in the columns
of the above matrices, therefore the rank is less than the number of
states. As consequence the outputs $h_1$ and $h_2$ make the systems
unobservable.

For $j=2$:
\begin{equation}
    \Xi_{b,2,1} = -\left[ \begin {array}{cc} {\left(-\frac{X_2}{X_1}\right)\frac {1}{{X_1}}}&\ \frac{1}{X_1}\\
    {\left(-\frac{X_2}{X_1}\right)\frac {2{X_2}}{{X_1}^2}}& {\frac {2 X_2}{{X_1}^{2}}}\end {array} \right]~, \nonumber
    \,\,\,
\Xi_{b,2,2} =     \left[ \begin {array}{cc} \frac{1}{X_2} & \left(-\frac {X_1}{X_2}\right) \frac{1}{X_2} \\
  \frac {2{X_1}\ c^2\  } {{X_2}^2} & \left(-\frac {X_1}{X_2}\right)  \frac {2{X_1}\ c^2\  } {{X_2}^2}\end {array} \right]~. \nonumber
\end{equation}

Again, it is possible to notice the linear dependence in the columns
of the above matrices, therefore the rank is less than the number of
states. As consequence the outputs $h_1$ and $h_2$ make the systems
unobservable.

However, for the outputs $h_3$ and $h_4$, i.e., if one can know
directly the zero modes or their derivatives, respectively, then

\begin{equation}
  \Xi_{b,1,3} = \left[ \begin {array}{cc} 1&0\\\noalign{\medskip}0&1\end {array} \right]~, \nonumber
\qquad
 \Xi_{b,1,4} =  \left[ \begin {array}{cc} 0&1\\\noalign{\medskip}+ c ^{2}&0\end {array} \right]~. \nonumber
\end{equation}

and

\begin{equation}
  \Xi_{b,2,3} = \left[ \begin {array}{cc} 1&0\\\noalign{\medskip}0&1\end {array} \right]~, \nonumber
\qquad
 \Xi_{b,2,4} =  \left[ \begin {array}{cc} 0&1\\\noalign{\medskip}- c ^{2}&0\end {array} \right]~. \nonumber
\end{equation}

for $j=1$ and $j=2$, respectively.

%\begin{eqnarray}
%   \Xi_{b,2,4} &=&  \left[ \begin {array}{cc} 0&1\\\noalign{\medskip}- c ^{2}&0\end {array} \right]~. \nonumber
%\end{eqnarray}
%
Both pairs of matrices have linear independent columns in an obvious
way and therefore the cosmological system is observable.

%.................................THE FERMIONIC CASE
\subsection{The Fermionic Case}

We write the cosmological dynamical system derived from
Eq.~(\ref{f1}) in the form
\begin{eqnarray} \Gamma _f: \ \ \ \left \{
\begin{array}{l}
    \frac{d{X}_1}{d\eta} = X_2\\
    \frac{d{X}_2}{d\eta} = -\ c\ F_j(\eta)\ X_1
    \end{array} \right.
\end{eqnarray}

where we defined
%...................
\begin{eqnarray}
    F_1(\eta) = c \left( 1+2\, %\left(
    \tan ^2 c \eta \right)\\  %\right) ^{2} \right) \\
    F_2(\eta) = c \left( -1+2\, \coth ^2  c \eta  \right) %{2} \right) %\\
   % c &=& \frac{{3}\ {\gamma}}{2}-1
\end{eqnarray}
%....................
In matrix form
\begin{eqnarray}
    \frac{dX}{d\eta} = f_j(X)~,
\end{eqnarray}
where $X \in \mathbf{R}^2$. Using the same outputs as in the bosonic
case and defining the codistribution in the same way we get

\begin{equation}
    \Xi _{f,1,1} = -\left[ \begin {array}{cc} {\left(-\frac{X_2}{X_1}\right)\frac {1}{{X_1}}}&\ \frac{1}{X_1}\\
    {\left(-\frac{X_2}{X_1}\right)\frac {2{X_2}}{{X_1}^2}}& {\frac {2 X_2}{{X_1}^{2}}}\end {array} \right]~, \nonumber
    \,\,\,
\Xi _{f,1,2} =  c\left[ \begin {array}{cc} \frac{1}{X_2} & \left(-\frac {X_1}{X_2}\right) \frac{1}{X_2} \\
  \frac {2{X_1}\ F_1(\eta) } {{X_2}^2} & \left(-\frac {X_1}{X_2}\right)  \frac {2{X_1}\ F_1(\eta) } {{X_2}^2}\end {array} \right]~. \nonumber
\end{equation}

The linear dependence in the columns of the above matrices is
manifest and therefore the rank is less than the number of states.
We conclude that the outputs $h_1$ and $h_2$ make the system
unobservable.

\medskip

Using now $F_2(\eta)$, the matrices

\begin{equation}
    \Xi _{f,2,1} = -\left[ \begin {array}{cc} {\left(-\frac{X_2}{X_1}\right)\frac {1}{{X_1}}}&\ \frac{1}{X_1}\\
    {\left(-\frac{X_2}{X_1}\right)\frac {2{X_2}}{{X_1}^2}}& {\frac {2 X_2}{{X_1}^{2}}}\end {array} \right]~, \nonumber
    \,\,\,
\Xi _{f,2,2} =     c\left[ \begin {array}{cc} \frac{1}{X_2} & \left(-\frac {X_1}{X_2}\right) \frac{1}{X_2} \\
  \frac {2{X_1}\ F_2(\eta) } {{X_2}^2} & \left(-\frac {X_1}{X_2}\right)  \frac {2{X_1}\ F_2(\eta) } {{X_2}^2}\end {array} \right] \nonumber
\end{equation}
%....................
are linear dependent at the level of their columns, so the rank is
less than the number of states and the outputs $h_1$ and $h_2$ make
the cosmological systems unobservable.

However, again, if one can know directly the zero modes or their
derivatives, then
\begin{equation}
  \Xi_{f,1,3} = \left[ \begin {array}{cc} 1&0\\\noalign{\medskip}0&1\end {array} \right]~, \nonumber
  \qquad
\Xi_{f,1,4} =  \left[ \begin {array}{cc} 0&1\\\noalign{\medskip}- c
F_1(\eta)&0\end {array} \right] \nonumber
\end{equation}

and

\begin{equation}
  \Xi_{f,2,3} = \left[ \begin {array}{cc} 1&0\\\noalign{\medskip}0&1\end {array} \right]~, \nonumber
  \qquad
\Xi_{f,2,4} =  \left[ \begin {array}{cc} 0&1\\\noalign{\medskip}- c
F_2(\eta)&0\end {array} \right]~. \nonumber
\end{equation}

%\begin{eqnarray}
 %  \Xi_{f,2,4} &=&  \left[ \begin {array}{cc} 0&1\\\noalign{\medskip}- c F_1(t)&0\end {array} \right]~. \nonumber
%\end{eqnarray}
%
All these four matrices have linear independent columns and
therefore the cosmological system is observable.

%.................................................................................................................CONCLUSION
%\section{Conclusion}

%The simple but rigorous techniques of mathematical nonlinear
%observability analysis show that the FRW barotropic zero modes
%cannot be determined if the measured quantity is the Hubble
%parameter alone or its inverse. This implies that the dynamical FRW
%barotropic state(s) remain undetermined if only the Hubble parameter
%is known. In other words, the usual type of cosmological
%measurements relying on the Hubble parameter are not the ones
%providing the complete dynamical information about the universe. On
%the other hand, having the chance to know the FRW zero modes gives
%us the opportunity to know dynamically the whole system. This,
%indirectly, offers the possibility that by fitting one can infer the
%initial conditions of the FRW universe.

%to the precise definition of mathematical
%observability of barotropic universes.

\section{Nonlinear Observability: Definitions and \\Basic Theorem of
Observability}

%\subsection*{Linear Observability.} Consider the linear system
%\begin{eqnarray}
%\dot{X} &=& AX(t)+Bu(t) \label{lineal1} \\
%y &=& CX(t)+Du(t) \label{lineal2}
%\end{eqnarray}
%Then one can introduce:

%\begin{definition}[Linear Systems Observability]\cite{Wolovich} and
%\cite{Donald}\\
%The linear state system given by equations \ref{lineal1} and
%\ref{lineal2} is called \emph{observable} if for any unknown initial
%state $X(0)$, there exist a finite $t_1>0$ such that the knowledge
%of the input $u$ and the output $y$ over $[0,t_1]$ suffices to
%determine uniquely the initial state $X(0)$. Otherwise, the system
%is said to be unobservable.
%\end{definition}

%\subsection*{Nonlinear Observability}

%Frequently, the following form is used in controlled systems

%\begin{eqnarray} \Gamma: \ \ \ \left \{
%\begin{array}{l}
%\dot{X}=\mathbf{f}(X)+\mathbf{g}(X)\mathbf{u} \label{afin en u}
%\\\,\,y=h(X) \label{salida}
%\end{array}
%\right.
%\end{eqnarray}

%\noindent where $\mathbf{u} \in \Omega \subset R^{1}$, $X \in
%\mathcal{S} \subset R^{2}$ and $\mathbf{f}$ and $\mathbf{g}$ are
%$\mathcal{C}^\infty $ functions. Note that in general
%$f(X)=\mathbf{f}(x)+\mathbf{g}(X)\mathbf{u}$. For the sake of
%simplicity we denote by $\Gamma$ the set of equations \ref{afin en
%u}.% and \ref{salida}.

The strictly mathematical material of this section refers to the
dynamical system of equations $\Gamma$ given in Eq.~(\ref{salida}).

\begin{Definition} %This is a text of a definition.
%$$ax + by + c = 0.$$
%\begin{definition}
[Indistinguishability and Observability] \cite{Hermann 77} \label{Def.
Indistinguibilidad}\\
Two states $X_a$, $X_b \in \mathcal{S}$ are said to be
indistinguishable (denoted by $X_a\texttt{I}X_b$) if $(\Gamma,X_a)$
and $(\Gamma,X_b)$ realize the same input-output map, i.e. for every
admissible input $u(t) \in \mathbf{u}$ defined in $[t_o,t_1]$
\begin{eqnarray}
\Gamma_{X_a}(u(t))=\Gamma_{X_b}(u(t)) \nonumber
\end{eqnarray}
Let $\texttt{I} \left ( X_o \right ) $ be the set of points
indistinguishable from $X_o$. The system $\Gamma$ is said to be
observable at $X_o$ if $\texttt{I} \left ( X_o \right ) = \left \{
X_o \right \} $ and is observable if $\texttt{I} \left ( X \right )
= \left \{ X \right \} $ for every $X \in \mathcal{S}$
%\end{definition}
\end{Definition}

Since the observability is a global concept distinguishing between
all the points in $\mathcal{S}$ is a difficult goal. Therefore, we
need a local concept and furthermore we merely need to distinguish
$X_o$ from its neighbors. For this purpose we define the concept of
local weak observability.

\begin{Definition}[Local Weak Observability]\label{DEF Obervabilidad Alineal}\cite{Hermann
77}\\ We shall say that $\Gamma$ is locally weakly observable at
$X_o$ if there exist an open neighborhood $\verb"U"$ of $X_o$ such
that for every open neighborhood $\verb"V"$ of $X_o$ contained in
$\verb"U"$, $\texttt{I}_{\verb"V"} \left ( X_o \right ) = \left \{
X_o \right \} $ and is locally weakly observable if
$\texttt{I}_{\verb"V"} \left ( X \right ) = \left \{ X \right \} $
for every $X \in \mathcal{S}$.
\end{Definition}

The advantage of the {\em local weak observability} over the other
concepts is that its proof is just necessary a simple algebraic
test.

%...........................................................................................DEFINITION 4
\begin{Definition}[Lie Derivative of a Real Valued Function]\cite{Isidori}\label{Derivada de
Lie}\\ Consider a smooth vector field $f$ and a function $\lambda$,
defined in an open set $U \subset \mathbb{R}^n$. The Lie derivative
of $\lambda$ along $f$ is the function
$L_{f}\lambda:\mathbb{R}^n\longrightarrow \mathbb{R}$ defined by:
\begin{eqnarray}
L_{f}\lambda (X) =\frac{\partial \lambda}{\partial X}
f(X)=\sum_{i=1}^n \frac{\partial \lambda}{\partial X_i} f_i(X)~.
\nonumber
\end{eqnarray}
\end{Definition}

\begin{Definition}[Observability Smooth Codistribution]\label{Algoritmo Isidori Obser}$\,$\\
The observability smooth codistribution can be generated
incrementally in the Isidori algorithmic way \cite{Isidori} as
follows
\begin{eqnarray}
\Omega_0&=&span\{h(X)\}\nonumber\\
\Omega_k&=&\Omega_{k-1}+\sum_{i=1}^{q}span\left\{L_{\tau_i}\Omega_{k-1}\right\}~.\nonumber
\end{eqnarray}
This smooth codistribution is called the observability space
$\mathcal{O}$. $L_{\tau_i}\Omega$ is defined in {\bf Definition
\ref{Derivada de Lie}}.
\end{Definition}

\begin{Definition}[Differential of {\it f}]\label{gradiente}\cite{Isidori}\\
Let $f:\mathbb{R}^n\longrightarrow \mathbb{R}$ be a smooth function.
The differential of $f$ is defined by:
\begin{eqnarray}
\verb"d" f=\frac{\partial \ f}{\partial X}= \left [ \frac{\partial \
f}{\partial X_1} \ \ \frac{\partial \ f}{\partial X_2} \ \ \cdots \
\ \frac{\partial \ f}{\partial X_n} \right ]~. \nonumber
\end{eqnarray}
\end{Definition}

\begin{Definition}[Observability Rank Condition] \label{Obs rank condition
DEF} $\,$ \\ $\Gamma$ is said  to satisfy the observability rank
condition at $X_o$ if the dimension of $\verb"d"\mathcal{O}(X_o)$ is
$n$ where $n$ is the dimension of the system and the differential
$\emph{d}$ is given by {\bf Definition \ref{gradiente}}. Moreover,
$\verb"d"\mathcal{O}(X)$ satisfies the observability rank condition
if this is true for any $X \in \mathcal{S}$.
\end{Definition}

\begin{Theorem} [Basic Theorem of Observability]$\,$\\
If $\Gamma$ satisfies the observability rank condition in {\bf
Definition \ref{Obs rank condition DEF}} at $X_o$
%if the dimension of $\verb"d"\mathcal{O}(X_o)$ is complete,
then $\Gamma$ satisfies the observability rank condition for any $X
\in \mathcal{S}$ if the dimension of $\verb"d"\mathcal{O}(X)$ is
complete.
%for every
%$X \in \mathcal{S}$.
\end{Theorem}
%\begin{Proof}
The complete proof of this theorem is given in \cite{Hermann 77}.
%\end{Proof}

%..............................................................................

%\section{Main Results} These are the main results of the paper.

%\begin{Theorem} This is a text of a theorem.

%\begin{equation} ax^2 + bx + c = 0. \end{equation}

%\end{Theorem}

%\begin{Lemma} This is a text of a lemma.

%\end{Lemma}

%.................................................................................................................CONCLUSION
\section{Conclusion}

In this paper we have treated the FRW barotropic zero modes as
dynamical state(s) of the universe for which we applied the rigorous
techniques of nonlinear observability as defined in engineering by
Hermann and Krener, and further discussed by Isidori in his
textbook. We have shown that these zero modes remain undetermined if
only the Hubble parameter is known, which in general is an expected
result for cosmologists since it appears that the Hubble parameter
provides only partial though important cosmological information. In
other words, the usual type of cosmological measurements relying on
the Hubble parameter are not the ones providing the complete
dynamical information about the universe. On the other hand, we have
obtained the interesting result that having the chance to know
directly the FRW zero modes gives us the opportunity to know
dynamically the whole system. This, indirectly, offers the
possibility that by fitting one can infer the initial conditions of
the FRW universe.
%to the precise definition of mathematical
%observability of barotropic universes.

\bigskip

{\bf ACKNOWLEDGEMENTS.} \\%This is a text of acknowledgements.
The second author wishes to thank IPICyT for a fellowship that
allowed him to collaborate with the first author on the research of
this paper.

{\bf Received: August 31, 2006}

\end{document}